\documentclass[preprint,review]{elsarticle}

\usepackage{graphicx}
\usepackage{amsmath}
\usepackage{amssymb}
\usepackage{mathtools}
\usepackage{xspace}
\usepackage{lineno}
\usepackage{hyperref}
\usepackage{slashed}

\biboptions{sort&compress}

\newcommand{\sss}{\scriptscriptstyle}
\newcommand{\D}{\ensuremath{{^2\text{H}}}\xspace}
\newcommand{\Dreaction}[3]{\ensuremath{\D(\gamma,#1)#2#3}\xspace}
\newcommand{\Nreaction}[3]{\ensuremath{#1(\gamma,#2)#3}\xspace}
\newcommand{\KYN}{\Dreaction{K}{Y}{N}}
\newcommand{\KpYN}{\Dreaction{K^+}{Y}{N}}

\newcommand{\KY}{\Nreaction{N}{K}{Y}}

\newcommand{\KpL}{\Nreaction{p}{K^+}{\Lambda}}
\newcommand{\KpSO}{\Nreaction{p}{K^+}{\Sigma^0}}
\newcommand{\KOSp}{\Nreaction{p}{K^0}{\Sigma^+}}

\newcommand{\KpSm}{\Nreaction{n}{K^+}{\Sigma^-}}

\newcommand{\mD}{\ensuremath{m_D}}

\newcommand{\mN}{\ensuremath{m_N}}

\newcommand{\fvec}[1]{\ensuremath{\textit{\textbf{#1}}}} 
\newcommand{\vp}[2][]{\ensuremath{\vec{p}_{#2}^{#1}}} 

\newcommand{\fvp}[2][]{\ensuremath{\fvec{p}_{#2}^{#1}}}
\newcommand{\sfvp}[2][]{\ensuremath{\slashed{\fvec{p}}_{#2}^{#1}}}

\newcommand{\fvg}[1][]{\fvp[#1]{\gamma}}
\newcommand{\fvd}[1][]{\fvp[#1]{D}}
\newcommand{\fvk}[1][]{\fvp[#1]{K}}
\newcommand{\fvy}[1][]{\fvp[#1]{Y}}
\newcommand{\fvn}[1][]{\fvp[#1]{N}}
\newcommand{\fvt}[1][]{\fvp[#1]{T}}
\newcommand{\vg}[1][]{\vp[#1]{\gamma}}
\newcommand{\vd}[1][]{\vp[#1]{D}}
\newcommand{\vk}[1][]{\vp[#1]{K}}
\newcommand{\vy}[1][]{\vp[#1]{Y}}
\newcommand{\vn}[1][]{\vp[#1]{N}}

\newcommand{\bra}[1]{\left\langle #1\right|}
\newcommand{\ket}[1]{\left|#1\right\rangle}

\newcommand{\figref}[1]{Fig.~\ref{fig:#1}}
\renewcommand{\eqref}[1]{(\ref{eq:#1})}

\journal{}

\begin{document}

\begin{frontmatter}

\title{Kaon photoproduction from the deuteron in a Regge-plus-resonance approach}

\author{P.~Vancraeyveld}\ead{pieter.vancraeyveld@ugent.be}
\author{L.~De~Cruz}
\author{J.~Ryckebusch\corref{author}}\ead{jan.ryckebusch@ugent.be}
\author{T.~Vrancx}
\address{Department of Physics and Astronomy, Ghent University, Proeftuinstraat 86, B-9000 Gent, Belgium}
\cortext[author]{Corresponding author}

\begin{abstract}
  We present a Regge-inspired effective-Lagrangian framework for kaon
  photoproduction from the deuteron.  Quasi-free kaon production is
  investigated using the Regge-plus-resonance (RPR) elementary
  operator within the relativistic plane-wave impulse approximation.  
  The RPR model was developed to describe photoinduced and
  electroinduced charged-kaon production off protons.  We show how
  this elementary operator can be transformed in order to account for
  the production of neutral kaons from both protons and neutrons.  The
  model results for kaon photoproduction from the deuteron compare
  favourably to the \KYN data published to date.
\end{abstract}



\end{frontmatter}


\section{Introduction}
\label{sec:intro}

Electromagnetic production of strangeness plays a prominent role in
the quest to chart the excitation spectrum of the nucleon. Since the
production mechanism inevitably involves quark-antiquark components of
the nucleon's sea, the reaction has the potential to probe unexplored
aspects of the nucleon's structure.

Leading experimental facilities have contributed to a large database
for the \KpL and \KpSO observables
\cite{ResonanceReview,ReviewExperiments}.  The self-analysing weak
decay of hyperons $Y$ is an enormous asset, since it
facilitates the determination of the recoil 
polarisation.  Hence, a wide range of single- and double-polarisation
observables can be accessed by combining a polarised beam and/or
target.  This paves the way for the determination of a \emph{complete}
set of observables.

In addition to the study of $p(\gamma,K^+)Y$, it pays to consider strangeness
production on more complex targets, such as the deuteron.  First,
owing to the deuteron's weak binding, it is ideally suited as an
effective neutron target and gives access to the elementary
$n(\gamma,K)Y$ reaction process.  Second, by comparing reactions off
free and bound protons, our understanding of nuclear-medium effects is
put to the test.  An important source of medium effects are the
rescatterings between the hyperon, nucleon and kaon.  These can be
considered as an undesirable background effect when exploiting the
deuteron as a neutron target, as it obscures the physics at the
photon-neutron-kaon vertex.  On the other hand, the final-state
interactions~(FSI) provide us with a tool to improve
our understanding of the hyperon-nucleon ($YN$) and kaon-nucleon ($KN$) potentials.
Focusing on kinematic regions with major hyperon
rescatterings allows one to gain access to the elusive
$YN$ interaction.

This letter investigates semi-inclusive strangeness production from
the deuteron within the relativistic plane-wave impulse approximation
(RPWIA).  The next section introduces the Regge-plus-resonance~(RPR)
formalism for modelling elementary strangeness production.
Section~\ref{sec:deuteron} focuses on how to embed the RPR production
operator in the nuclear medium.  In Section~\ref{sec:results}, we
present model calculations and compare them to data. Here, the sensitivity
of the results to the model's assumptions is investigated. Finally, we
present our conclusions and indicate directions for future work.


\section{Elementary kaon production}
\label{sec:elementary}

A relatively high production threshold and the absence of plain
resonant structures in the energy dependence of the measured cross section
point towards a dominance of non-resonant
contributions to electromagnetic kaon production. This sets
strangeness production apart from reactions such as $N(\gamma,\pi)N$
and $N(\pi,\pi')N'$, and calls for a unique formalism that addresses
the \KY peculiarities.  The RPR approach seeks to decouple the
determination of the coupling constants for the background and the
resonant diagrams.  This results in a hybrid model which accounts for
electromagnetic kaon production from threshold up to
$E_{\gamma}=16\;\text{GeV}$.  The generic structure of the transition
current operator in the RPR approach reads
\begin{subequations}
\begin{align}
\hat{\bf J}_{KY}
&= \hat{\bf J}^{\sss K^{+}(494)}_{\text{Regge}} \label{eq:rpr_K}\\
&\quad+ \hat{\bf J}^{\sss K^{\ast+}(892)}_{\text{Regge}} \label{eq:rpr_Kstar}\\
&\quad+ \hat{\bf J}^{\sss\text{$p$,elec}}_{\text{Feyn}} \times
\mathcal{P}^{\sss K^{+}(494)}_{\text{Regge}} \times
\left( t-m_{\sss K^{+}}^2 \right) \label{eq:rpr_Selec}\\
&\quad+ \sum_{\sss N^{\ast}} \hat{\bf J}^{N^{\ast}}_{\text{Feyn}}
+ \sum_{\sss \Delta^{\ast}} \hat{\bf J}^{\Delta^{\ast}}_{\text{Feyn}}\,. \label{eq:rpr_res}
\end{align}
\end{subequations}
The non-resonant contributions of Eqs.~\eqref{rpr_K}
and~\eqref{rpr_Kstar} are efficiently modelled in terms of $t$-channel
$K^+(494)$ and $K^{\ast+}(892)$ Regge-trajectory exchange
\cite{GuidalPhotoProdKandPi}.  The three coupling constants can be
determined from the high-energy ($E_{\gamma}\gtrsim 4\;\text{GeV}$)
data~\cite{RPRbayes}.  A crucial constraint is gauge invariance.  The
$t$-channel Born diagram of Eq.~\eqref{rpr_K} by itself is not gauge invariant.
Adding the electric part of a Reggeized $s$-channel
Born diagram~\eqref{rpr_Selec} ensures that the $p(\gamma,K^+)Y$
amplitude is gauge invariant \cite{GuidalPhotoProdKandPi}.  The Regge
amplitudes are supplemented with $s$-channel
nucleon ($N^\ast$) and delta ($\Delta^\ast$) resonance-exchange
diagrams~\eqref{rpr_res}, whose parameters are optimised to data in
the resonance-region ($E_{\gamma}\lesssim4\;\text{GeV}$) while keeping
the background anchored.

High-quality kaon-photoproduction data over an extended energy and
angular range is only available for \KpL and \KpSO.  An economical
description of these reactions has been
obtained~\cite{RPRlambda,RPRsigma,RPRelectro,RPRneutron} and will be
referred to as RPR-2007\footnote{%
The RPR-2007 model corresponds to the 
$p(\gamma,K^+)\Lambda$ model labelled `RPR-2 + $D_{13}(1900)$'
and the $p(\gamma,K^+)\Sigma^0$ model labelled as `RPR-$3^\prime$'
in Table~1 of Ref.~\cite{RPRelectro}.
The relevant coupling constants are listed in Appendix~I of Ref.~\cite{PhdPieter}.
}.  For $p(\gamma,K^+)\Lambda$, a
set of established nucleon resonances turns out to be insufficient.
The addition of a $D_{13}(1900)$ resonance makes it possible to
accurately describe both photo- and electroproduction
data~\cite{RPRlambda,RPRelectro}.  The
$p(\gamma^{(\ast)},K^+)\Sigma^0$ data, on the other hand, is properly
described within the RPR framework considering established
$N^\ast$'s and $\Delta^\ast$'s~\cite{RPRsigma,RPRelectro}.

A total of six $N(\gamma,K)Y$ reactions can be treated within a single
theoretical framework.  The two $\Lambda$ and the four $\Sigma$
reactions can be described by a single set of parameters based on the
RPR-2007 model which is optimised to \KpL and \KpSO data.  The
reaction channels can be interrelated by converting the coupling
constants which feature in the interaction Lagrangians while
maintaining gauge invariance.  For $K^0\Lambda$ and $K^0\Sigma^0$ production, the
kaon-exchange amplitude~\eqref{rpr_K} vanishes and gauge-invariance
restoration becomes irrelevant.  Further, $n(\gamma,K^+)\Sigma^-$ is
the only channel with a neutron target and a charged
kaon. Accordingly, the electric part of the $s$-channel Born
diagram~\eqref{rpr_Selec} is identically zero.  A gauge-invariant
amplitude is obtained by including the electric part of a Reggeized
$u$-channel Born diagram.

In the strong-interaction vertex, one can fall back on SU(2) isospin
symmetry to find the relevant conversion factors, since the hadronic
couplings are proportional to the Clebsch-Gordan
coefficients~\cite{RPRneutron,PhdPieter}.
We adopt the isospin conventions of Ref.~\cite{RPRneutron}. 
\begin{subequations}\label{eq:isospinRelations}
  The strong coupling constants for the $\Lambda$-production channels
  are isospin independent.
The strong-interaction vertices for $p(\gamma,K^0)\Sigma^+$ can be related
to those for $p(\gamma,K^+)\Sigma^0$ 
\begin{equation}\begin{split}
 g_{K^{(\ast)0}\Sigma^+p} &= \sqrt{2}\,g_{K^{(\ast)+}\Sigma^0p}\, ,\\
 g_{K^{(\ast)0}\Sigma^+N^{\ast+}} &= \sqrt{2}\,g_{K^{(\ast)+}\Sigma^0N^{\ast+}}\, ,\\
 g_{K^{(\ast)0}\Sigma^+\Delta^{\ast+}} &= 
\dfrac{-1}{\sqrt{2}}g_{K^{(\ast)+}\Sigma^0\Delta^{\ast+}} \, .
\end{split}\end{equation}
Similar expressions apply for $n(\gamma,K^+)\Sigma^-$
\begin{equation}\begin{split}
 g_{K^{(\ast)+}\Sigma^-n} &= \sqrt{2}\,g_{K^{(\ast)+}\Sigma^0p}\, ,\\
 g_{K^{(\ast)+}\Sigma^-N^{\ast0}} &= \sqrt{2}\,g_{K^{(\ast)+}\Sigma^0N^{\ast+}}\, ,\\
 g_{K^{(\ast)+}\Sigma^-\Delta^{\ast0}} &= \dfrac{1}{\sqrt{2}}g_{K^{(\ast)+}\Sigma^0\Delta^{\ast+}}\, .
\end{split}\end{equation}
The transformation of the $p(\gamma,K^+)\Sigma^0$ amplitude to 
the $n(\gamma,K^0)\Sigma^0$ one, 
requires sign changes
\begin{equation}\begin{split}
 g_{K^{(\ast)0}\Sigma^0n} &= -\,g_{K^{(\ast)+}\Sigma^0p}\, ,\\
 g_{K^{(\ast)0}\Sigma^0N^{\ast0}} &= -\,g_{K^{(\ast)+}\Sigma^0N^{\ast+}}\, ,\\
 g_{K^{(\ast)0}\Sigma^0\Delta^{\ast0}} &= g_{K^{(\ast)+}\Sigma^0\Delta^{\ast+}}\, .
\end{split}\end{equation}
\end{subequations}

Unlike the coupling constants in the strong-interaction vertex, the
transformation of those in the electromagnetic-interaction vertex
cannot proceed without experimental input.  We first focus on reactions
with a neutron target and summarise some issues discussed in more detail
in Ref.~\cite{RPRneutron}.

The partial decay width for the radiative decay of a
resonance 
to the ground-state nucleon can be expressed in terms of photocoupling
helicity amplitudes $\mathcal{A}^N_{J}$ which can be linked with
current matrix elements.
The latter can, for example, be calculated within a quark
model~\cite{BonnEMff}, or
with phenomenological interaction Lagrangians.  Thereby, the
$N^{\ast}$~($\kappa_{N^*N}^{(1,2)}$) and
$\Delta^{\ast}$~($\kappa_{\Delta^\ast N}^{(1,2)}$) transition moments
can be related to the 
$\mathcal{A}^N_{J}$.  Inverting these relations and neglecting the
small proton-neutron mass difference, we find~\cite{RPRneutron}
\begin{subequations}\label{eq:EMratio}
\begin{align}
\frac{\kappa_{ {
N^*}n}}{ \kappa_{ { N^*}p}} &= \frac{\mathcal{A}^n_{1/2}}{\mathcal{A}^p_{1/2}} \;,\\
\intertext{for spin-1/2 resonances, and}
\frac{\kappa_{ {
N^*}n}^{\left( 1 \right)}}{ \kappa_{ { N^*}p}^{\left( 1 \right)}} &=
\frac{ \sqrt{3} \mathcal{A}^n_{1/2} \pm \mathcal{A}^n_{3/2}}{\sqrt{3} \mathcal{A}^p_{1/2} \pm
\mathcal{A}^p_{3/2}} \;, \\
\frac{\kappa_{ { N^*}n}^{\left( 2 \right)}}{ \kappa_{ {
N^*}p}^{\left( 2 \right)}} &= \frac{ \sqrt{3} \mathcal{A}^n_{1/2} -
\frac{m_p}{m_{ N^*}} \mathcal{A}^n_{3/2}}{\sqrt{3} \mathcal{A}^p_{1/2} -
\frac{m_p}{m_{ N^*}} \mathcal{A}^p_{3/2}} \;,
\end{align}
\end{subequations}
for spin-3/2 resonances.

As motivated in Ref.~\cite{RPRneutron}, we employ helicity amplitudes
of the SAID analysis SM95~\cite{SAID96}.  Table~\ref{tab:helamp}
lists the conversion factors for the resonances relevant to RPR-2007.
Obviously the ratios have considerable error bars.  Moreover, no
information for the $D_{13}(1900)$ and $P_{13}(1900)$ is available.
Therefore, we allow the ratios of the magnetic transition moments,
$\kappa_{ { N^*}n}^{\left( 1,2 \right)}/ \kappa_{ {N^*}p}^{\left( 1,2
  \right)}$, to vary between $-2$ and $+2$.  Since the transition
strengths for $\Delta ^{*} \rightarrow n\gamma,p\gamma$ are identical, we
include the $D_{33}(1700)$, $S_{31}(1900)$, $P_{31}(1910)$, and
$P_{33}(1920)$ with the electromagnetic coupling constants
determined from $p(\gamma,K^+)\Sigma^0$.

We evaluate the predictive power of RPR-2007 for kaon photoproduction
from the neutron
using the two available \KpSm data sets.  The LEPS results
~\cite{LEPSiso6} comprise differential cross sections (d.c.s.) and
photon-beam asymmetries for $\cos\theta_{K}^{\ast}\geq
0.65$ and $1.5\,\text{GeV}\leq
E_{\gamma}\leq2.4\,\text{GeV}$.  Recently, the CLAS collaboration has published 
a large set of $n(\gamma,K^+)\Sigma^-$
d.c.s.\ \cite{AnefalosIso6}.  These data cover 
incident photons from $0.8\,\text{GeV}$ to
$3.6\,\text{GeV}$ on a liquid-deuterium target.

\begin{table}
\caption{%
The ratio of the electromagnetic coupling constants to proton and neutron for selected nucleon resonances 
obtained with Eq.~(\ref{eq:EMratio}). 
The listed values are obtained using photocoupling helicity amplitudes from SAID analysis SM95~\cite{SAID96}.
No experimental information exists for resonances of mass $1900\,\text{MeV}$,
therefore we consider a broad range.%
}
\label{tab:helamp}
\begin{tabular}{lr@{$\,\pm\,$}rr@{$\,\pm\,$}rr@{$\,\pm\,$}r}
\hline\noalign{\smallskip}
Resonance  & \multicolumn{2}{c}{$\frac{\kappa_{ {N^*}n}}{ \kappa_{ { N^*}p}}$} 
           & \multicolumn{2}{c}{$\frac{\kappa_{ {N^*}n}^{\left( 1 \right)}}{ \kappa_{ { N^*}p}^{\left( 1 \right)}}$}
           & \multicolumn{2}{c}{$\frac{\kappa_{ { N^*}n}^{\left( 2 \right)}}{ \kappa_{ {N^*}p}^{\left( 2 \right)}}$} \\ 
\noalign{\smallskip}\hline\noalign{\smallskip}
$S_{11}(1650)$ &$-0.22$ &$0.07$ &\multicolumn{2}{c}{$-$} &\multicolumn{2}{c}{$-$}  \\ 
$P_{11}(1710)$ &$-0.29$ &$2.23$ &\multicolumn{2}{c}{$-$} &\multicolumn{2}{c}{$-$} \\ 
$P_{13}(1720)$ &\multicolumn{2}{c}{$-$} &$-0.38$ &$2.00$ &$-0.50$ &$1.08$ \\ 
$D_{13}(1900)$ &\multicolumn{2}{c}{$-$} &$ 0.00$ &$2.00$ &$ 0.00$ &$2.00$ \\ 
$P_{13}(1900)$ &\multicolumn{2}{c}{$-$} &$ 0.00$ &$2.00$ &$ 0.00$ &$2.00$ \\ 
\noalign{\smallskip}\hline
\end{tabular}
\end{table}

\begin{figure}
\centering
\includegraphics[scale=.59]{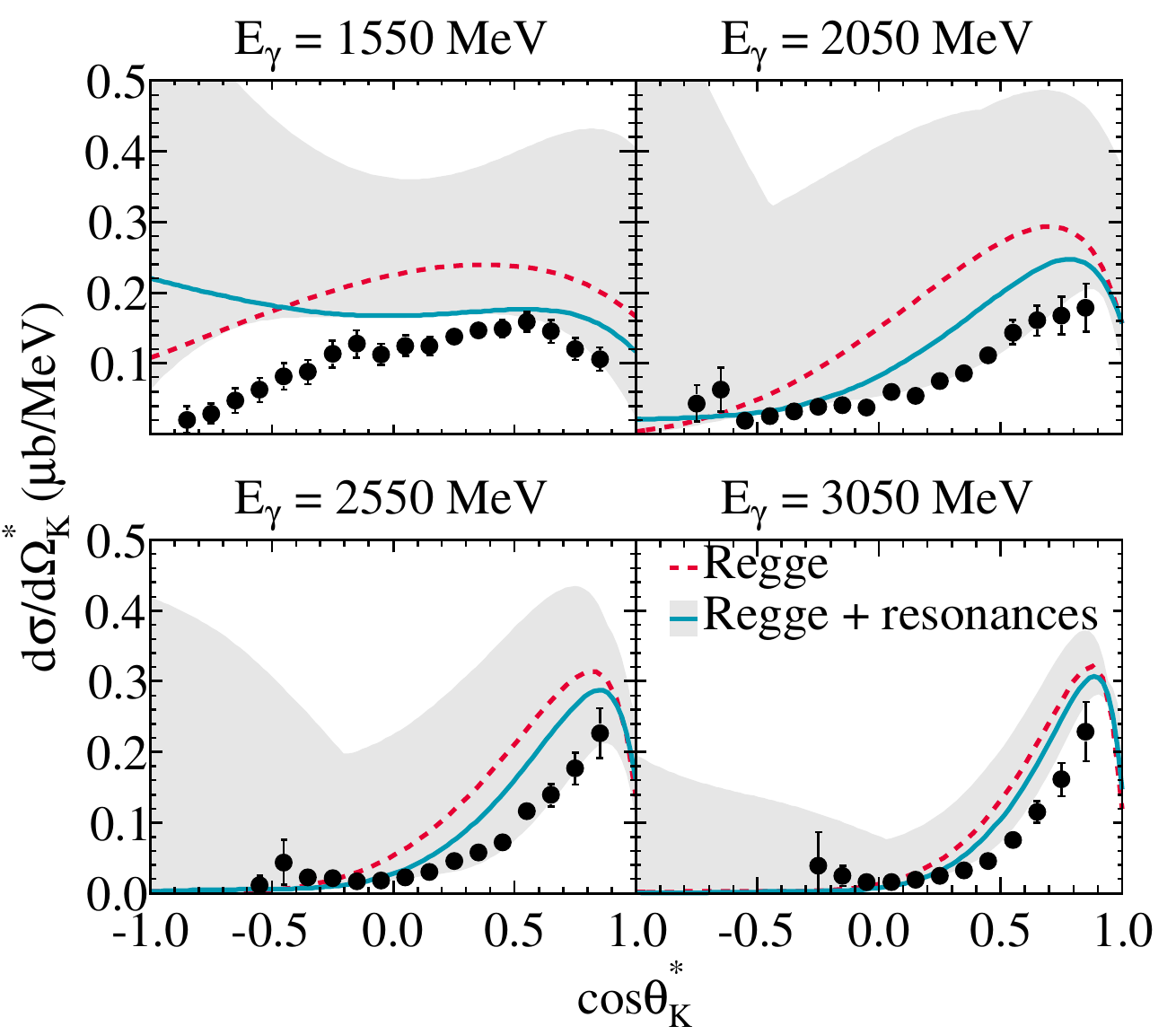}\caption{%
  The \KpSm differential cross section as a function of the kaon scattering angle
  $\cos\theta_K^\ast$ for 4 different values of the photon laboratory
  energy $E_\gamma$.  The dashed curve indicates the contribution of
  the Reggeized background, whereas the full curve corresponds to the
  complete RPR result, i.e.\ background and resonance contributions.
  The shaded area takes the uncertainties of the
  helicity amplitudes from Table~\ref{tab:helamp} into account.  Data are from
  Ref.~\cite{AnefalosIso6}. }
\label{fig:iso6}
\end{figure}

In \figref{iso6}, model predictions are set against CLAS results at
four values for $E_{\gamma}$.  At the highest energies, where
resonance exchange has nearly died out, the d.c.s.\ peaks at forward
angles.  The exponential decrease of the d.c.s.\ as a function of $\cos
\theta ^{*} _{K}$ is distinctive for Regge-trajectory exchange.  This
characteristic feature of the $n(\gamma,K^+)\Sigma^-$
d.c.s.\ disappears as one moves into the resonance region.  The quality
of agreement of the three-parameter Regge model is fair as the shape
of the cross section is reproduced, yet its strength is overestimated.
One observes destructive interference between the Reggeized background
and the resonances.  This effect
considerably improves the overall agreement with the d.c.s.\ data,
except at the lowest photon energy where the backward strength is
overpredicted.

The $N^\ast$ coupling constants, listed in Table~\ref{tab:helamp},
have considerable error bars which induce uncertainties.  Their impact
is assessed in \figref{iso6}, and turns out to be quite dramatic.  The
shaded area indicates the range of values obtained for
$d\sigma/d\Omega$ when the conversion factors of Table~\ref{tab:helamp} are varied within their
error bars.  The experimental ambiguities of the transformed photon
couplings result in deviations up to $100\,\%$ for the d.c.s.
Our results show that the resonant and Reggeized background
amplitudes interfere destructively.
This destructive interference is
affected by varying the electromagnetic coupling constants of the
resonances within their allowed ranges.  With a large conversion
factor, a particular resonance dominates and the cross section is
enlarged.
Hence, the
error band in \figref{iso6} is not positioned symmetrically around the
central RPR prediction.


When transforming the electromagnetic vertex for $N ( \gamma, K ^{0} )
Y$, the relevant coupling constants for $t$-channel $K^+(494)$ and
$K^{\ast+}(892)$ exchange are the charge of the kaon and the magnetic
transition moment $\kappa_{K^{\ast}K}$.  In the transformation from
$p(\gamma,K^+)Y^0$ to $N(\gamma,K^0)Y$, the electromagnetic vertices
of the resonance-exchange terms~\eqref{rpr_res} are unaffected.

In the $N(\gamma,K^0)Y$ channels,
the contributions from the kaon-exchange diagram~\eqref{rpr_K}
and the accompanying gauge-invariance-restoring $s$-channel diagram~\eqref{rpr_Selec} vanish.
Therefore, 
the $K^{\ast0}(892)$-exchange diagram~\eqref{rpr_Kstar} is the sole non-resonant contribution that survives.
The decay width of $K^\ast$ vector meson to the ground-state kaon can be directly linked to the square of the magnetic transition moment. 
Adopting the decay widths listed in the Review of Particle Physics~\cite{pdg},
one finds
\begin{equation}\label{eq:strongRelations}
 \dfrac{\kappa_{K^{\ast0}(892)K^{0}(494)}}{\kappa_{K^{\ast+}(892)K^{+}(494)}} = -1.53 \pm 0.10\,.
\end{equation}
The relative sign for these coupling constants cannot be deduced from
experiment.  We adopt the sign predicted by the quark models 
of Singer and Miller~\cite{SingerMiller} and the Bonn group~\cite{VanCauteren:2005sm}.

The \KOSp reaction is the only neutral-kaon production channel where
data is available.  The use of Eqs.~\eqref{isospinRelations}
and~\eqref{strongRelations} implies that the \KOSp yield will be
roughly $4.7$ times larger than the \KpSO one.  

The data~\cite{RPRsigma,PhdPieter}, however, indicate that the \KOSp
yield is about 50\% of the \KpSO yield near
threshold.  The absence of $K^0\Sigma^+$ data at high-energies makes it
impossible to constrain possible additional contributions to the Regge
amplitude~\cite{PhdPieter}. Therefore, we adopt a pragmatic
approach, and fit the electromagnetic coupling constant of the
$K^{\ast}(892)$ Regge trajectory to the available \KOSp data to
find~\cite{PhdPieter}
\begin{equation}\label{eq:ratioFit_KSP2KS0}
  \dfrac{\kappa_{K^{\ast0}(892)K^{0}(494)}}{\kappa_{K^{\ast+}(892)K^{+}(494)}} 
  = 0.054 \pm 0.010\,.
\end{equation}
All other parameters in RPR-2007 are kept fixed. 
Despite the fact that only one free parameter is introduced, 
we attain $\chi^2/N_{\text{data}}=3.39$.
With the fitted ratio of Eq.~\eqref{ratioFit_KSP2KS0} 
the $K^{\ast}(892)$ trajectory is strongly suppressed
and resonance exchanges play a more dominant role in \KOSp.


\section{Modelling the \texorpdfstring{\KYN}{2H(g,K)YN} reaction}
\label{sec:deuteron}

Having established a model for all $N(\gamma,K)Y$ channels, they can now
be embedded in the nuclear medium.  The dominant contribution to 
\KYN
stems from the quasi-free process: the
photon interacts with a single bound nucleon and produces the strange
meson and hyperon.  All final-state particles subsequently leave the
interaction region as plane waves. All of them, however, 
can undergo elastic and inelastic rescatterings before reaching the
detectors.  These FSI can be considered a curse.  Yet, FSI also
present unique opportunities to access the $YN$ interaction.

Yamamura~\mbox{\textit{et al.}} pioneered \KYN investigations with a modern
elementary kaon-production operator~\cite{Yamamura,Miyagawa2006},
and included the $YN$ FSI with the
Nijmegen $YN$ potential. The approach was extended \cite{Salam2004} to
include two-step production and kaon-nucleon rescattering.  Adopting
this model, neutral-kaon photoproduction has been studied.
Refs.~\cite{Salam2006,Salam2009} focus on the extraction of the
elementary amplitude.  A different study on the influence of $YN$
rescattering using the $P$-matrix approach is presented in
Ref.~\cite{Kerbikov2000}.  Maxwell considered a host of rescattering
diagrams with $\pi$, $\eta$ and $K$ exchanges between the active and
the spectator nucleon~\cite{Maxwell2004a,Maxwell2004}.  In
Refs.~\cite{Bydzovsky2010a,Bydzovsky2010b}, \KYN is investigated in
the non-relativistic plane-wave impulse approximation~(NRPWIA) with various isobar models,
demonstrating the importance of
a reliable elementary-production operator.  Gasparyan~\mbox{\textit{et
    al.}} studied the possibility of extracting the low-energy $\Lambda n$
scattering parameters~\cite{Gasparyan2007}.  Laget identified
well-defined regions in phase space where $KN$ and $YN$ rescattering
dominate while the elementary amplitude is on shell and the momentum
of the spectator nucleon is low~\cite{Laget2006,Laget2006pentaquark}.

In our notation, the four-vectors of the photon, deuteron, kaon,
hyperon and nucleon are $\fvg(E_\gamma,\vg)$, $\fvd(E_D,\vd)$,
\fvk$(E_K,\vk)$, \fvy$(E_Y,\vy)$ and \fvn$(E_N,\vn)$.
In what follows, all variables will be expressed in the
laboratory frame, unless stated differently.

For the three-body final state, one can define a large number of
coordinate systems.  We label the final state as $1+(23)$.  The
coordinate system $(x,y,z)$, in which we describe the reaction,
has its $z$ axis along \vg.  The $y$ axis is chosen
perpendicular to the plane spanned by the photon and particle $1$.

In order to improve on the statistics, the \KYN data are often
presented as semi-inclusive observables. Thereby, one integrates over
the phase space of one or two final particles.
We find for the semi-inclusive unpolarised d.c.s.\ 
\begin{multline}\label{eq:diffXsectionSemiInclusive}
  \frac{d^3\sigma_{\text{unpol}}}{d|\vp{1}|d\Omega_1}
  = \frac{1}{32(2\pi)^5}\frac{|\vp{1}|^2|\vp[\,\ast]{2}|}{\mD E_{\gamma} E_1 W_{23}}
  \\\times
  \frac{1}{3}
  \int d\Omega_2^* 
  \sum_{\lambda_D\lambda_N\lambda_Y}
  |\mathcal{T}_{\lambda_N,\lambda_Y}^{\lambda_D,\lambda_{\gamma}=+1}|^2 \, .
\end{multline}
Variables marked with an $*$ are evaluated in the c.m. frame of
particles 2 and 3.  Upon detecting the polarisation of the hyperon, a
recoil polarisation asymmetry $\Pi_y$ can be defined~\cite{PhdPieter}
\begin{multline}\label{eq:recpol}%
\Pi_y
\frac{d^3\sigma_{\text{unpol}}}{d|\vp{Y}|d\Omega_Y}
= \frac{1}{32(2\pi)^5}\frac{|\vp{Y}|^2|\vp[\,\ast]{N}|}{\mD E_{\gamma} E_Y W_{KN}}
\\\times
\frac{1}{3} \int d\Omega^{\ast}_N \sum_{\lambda_\gamma\lambda_D\lambda_N} 
\Im\left[
\left(\mathcal{T}_{\lambda_N,\lambda_Y=+\frac{1}{2}}^{\lambda_D,\lambda_{\gamma}}\right)^\ast
\mathcal{T}_{\lambda_N,\lambda_Y=-\frac{1}{2}}^{\lambda_D,\lambda_{\gamma}}
\right] \,.
\end{multline}

The dynamics of the reaction are contained in the transition amplitude
$\mathcal{T}_{\lambda_N,\lambda_Y}^{\lambda_D,\lambda_{\gamma}}$,
where $\lambda_{\gamma}$, $\lambda_D$, $\lambda_Y$ and $\lambda_N$
indicate the helicities,
which are defined in the laboratory frame. 

In this work, we adopt the impulse approximation, 
which states that the full nuclear many-body current operator can be approximated 
by a sum of one-body current operators.
In the \mbox{RPWIA},
the Lorentz-invariant transition amplitude is given by
\begin{multline}\label{eq:rpwia}%
  \mathcal{T}_{\lambda_N,\lambda_Y}^{\lambda_D,\lambda_{\gamma}}=
  \mp\sqrt{2}\;
  \overline{u}(\vp{Y},\lambda_Y)
  \epsilon^{\lambda_{\gamma}}_{\nu}\hat{\text{J}}^{\nu}_{KY} 
  \\\times
  \frac{m_T+\sfvp{T}}{m_T^2-\left(\fvt\right)^2}
  \xi^{\lambda_D}_{\mu}
  \Gamma_{\text{\tiny Dnp}}^\mu(\fvn,\fvd)
  \mathcal{C} \overline{u}^T(\vp{N},\lambda_N)\,,
\end{multline}
with $\mathcal{C}=-i\gamma^0\gamma^2$ the charge conjugation matrix.
The mass and four-vector of the struck nucleon are given by $m_T$ and $\fvt$.
The factor $\mp\sqrt{2}$ stems from isospin factors,
and the fact that the production operator acts on a single proton~(neutron)~\cite{PhdPieter}.
Since the spectator nucleon 
is on mass shell, 
the covariant $Dnp$-vertex $\bf\Gamma_{\text{\tiny Dnp}}$ is defined by~\cite{BlankenbeclerCook}
\begin{multline}
 \Gamma_{\text{\tiny Dnp}}^{\mu}(\fvn,\fvd) = 
  F(|\vec{p}|)\gamma^{\mu} - \frac{G(|\vec{p}|)}{m_N}p^{\mu}
\\
  -\frac{m_N-(\slashed{\fvec{p}}_D-\slashed{\fvec{p}}_N)}{m_N}
    \left( H(|\vec{p}|)\gamma^{\mu} - \frac{I(|\vec{p}|)}{m_N}p^{\mu} \right)\,,
\end{multline}
with $\fvec{p} = \frac{1}{2}\fvd-\fvn$.
The four scalar form factors $F$, $G$, $H$ and $I$ 
can be expressed in terms of 
the $S$-, $P$- and $D$-wave components of the deuteron wave function~\cite{BuckGross,PhdPieter}.

The target nucleon is obviously off its mass shell and the $\hat{\bf
  J}_{KY}$ of Eq.~\eqref{rpwia} is evaluated with one off-mass-shell
leg.  Owing to the deuteron's tiny binding energy, the virtuality is
minor for small spectator-nucleon momenta \vn. 
In order to assess the uncertainties induced by off-shell extrapolations,
we introduce the on-shell four-vector $\widetilde{\fvec{p}}_T
\stackrel{\text{\text{\tiny LAB}}}{\equiv}
\left(\sqrt{|\vn|^2+m_T^2},-\vn\right)$ of the target nucleon.  After
decomposing the nucleon propagator in Eq.~\eqref{rpwia}, we obtain for
the on-shell part of the RPWIA amplitude~\cite{PhdPieter}
\begin{multline}\label{eq:rpwia_onshell}%
  {\mathcal{T}}_{\lambda_N,\lambda_Y}^{\lambda_D,\lambda_{\gamma}}
  \stackrel{\text{\tiny LAB}}{=}
  \mp\sqrt{2}
  \sqrt{(2\pi)^3 2\mD}
  \frac{2E_N}{\mD}
  \\\times
  \sum_{\lambda_T}
  \bra{\fvk;\fvy,\lambda_Y}\hat{J}^{\lambda_{\gamma}}_{KY}\ket{\widetilde{\fvp{}}_T,\lambda_T}
  \Psi_{\lambda_T\lambda_N}^{++}(-\vp{N},\lambda_D)\,,
\end{multline}
with $\Psi^{++}$ the positive-energy deuteron wave function.  The
above equation represents the NRPWIA expression for the transition
amplitude, up to a kinematical factor $2E_N/\mD$.  The
amplitude of Eq.~\eqref{rpwia_onshell} is evaluated with all incoming
and outgoing particles on their mass shells.
It needs to be evaluated in the laboratory frame unlike Eq.~\eqref{rpwia}.


\section{Results}
\label{sec:results}

\begin{figure}
\centering
\includegraphics[width=.37\textwidth]{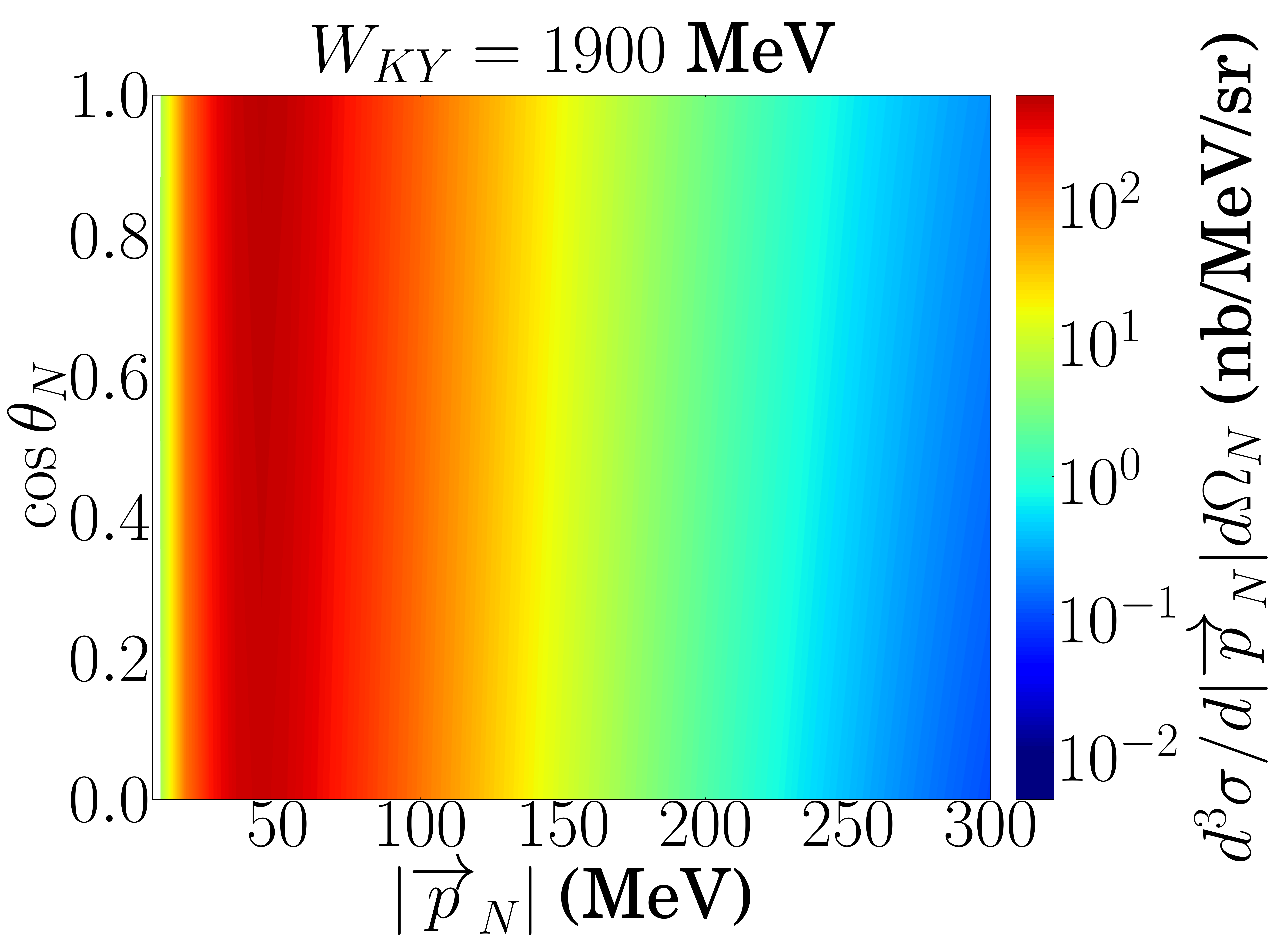}\caption{%
The semi-inclusive \KpYN differential cross section using the full RPR operator in the RPWIA
as a function of the nucleon momentum $|\vn|$ 
and scattering angle $\cos\theta_N$ in the LAB frame
at $W_{KY}=1900$ MeV.
}
\label{fig:missingMomentum}
\end{figure}

The \KYN results can be presented in terms of various combinations of the kinematic variables.
For small outgoing nucleon momenta ($|\vn|\ll\mN$) and vanishing FSI 
the d.c.s.\ can be approximated by~\cite{PhdPieter}
\begin{equation}\label{eq:specNapprox}
\frac{d^5\sigma}{d\vn d\Omega_{K}^{\ast}} 
\stackrel{\text{LAB}}{\approx}
\left( 1 + \frac{|\vn|}{E_N}\cos\theta_N\right)
\frac{ \rho_D(|\vn|) } { 2\pi}
\frac{d^2\sigma_{KY}}{d\Omega_{K}^{\ast}} \,,
\end{equation}
with $\rho_D$ the deuteron density and
$d^2\sigma_{KY}/d\Omega_{K}^{\ast}$ the elementary $KY$ cross section.
Owing to the factorised form~\eqref{specNapprox} of the d.c.s., known as the
non-relativistic spectator-nucleon approximation~\cite{LagetPhysRept},
we can assess which regions in phase space are to contribute most to
the reaction's strength.  The elementary amplitude
exhibits only mild variations with energy, whereas $\rho_D$ falls off
exponentially with increasing $|\vn|$.  Under those conditions
that the reaction is dominated by the RPWIA contribution, most
strength will emerge in phase-space regions corresponding to low $|\vn|$.
This is confirmed in \figref{missingMomentum},
where the semi-inclusive d.c.s.\ 
is shown as function of $|\vn|$ and $\cos\theta_N$.  We
notice an exponential falloff as $|\vn|$ increases, reminiscent of the
deuteron density.  The d.c.s.\ is nearly isotropic.
The bands in \figref{missingMomentum} are only slightly
tilted in the clockwise direction.  This indicates a mild dependence
on $\cos\theta_N$ that becomes stronger as the spectator momentum
rises.

\begin{figure}
\centering
\includegraphics[scale=.59]{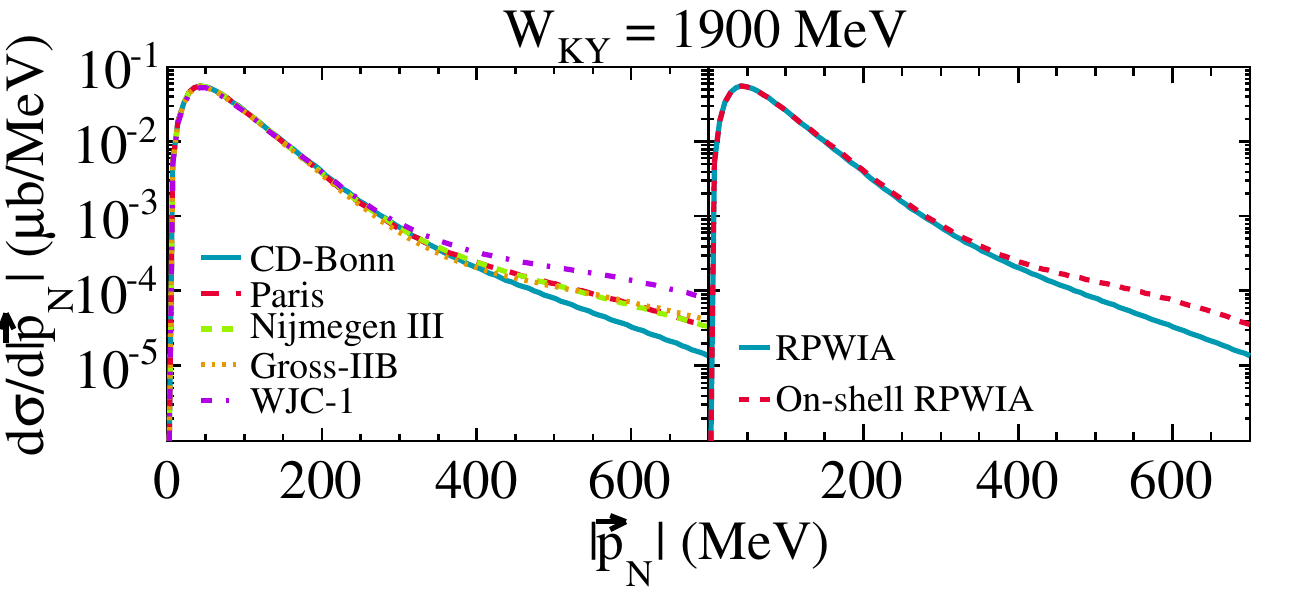}\caption{%
  The $^2\text{H}(\gamma,n)K^+\Lambda$ differential cross section as a
  function of the nucleon momentum $|\vn|$ at $W_{KY} =
  1900\,\text{MeV}$.  In the left panel, results are obtained within
  the RPWIA adopting different versions of the deuteron wave function:
  CD-Bonn~\cite{CDBonnWF}, Paris~\cite{ParisPot},
  Nijmegen-III~\cite{NijmegenPotential}, Gross-IIB~\cite{GrossWF} and
  WJC-1~\cite{GrossWJC}.  The right panel compares for the CD-Bonn
  wave function, the RPWIA result (solid curve) with the one obtained
  with the on-shell reduction of Eq.~\eqref{rpwia_onshell} (dashed
  curve).}
\label{fig:sensitivity}
\end{figure}

We investigate the sensitivity of the computed cross sections to the
various model ingredients in \figref{sensitivity}.  The role of
off-shell effects can be examined by comparing the RPWIA
of Eq.~\eqref{rpwia} to its on-shell reduction~\eqref{rpwia_onshell}.
In the right panel of
\figref{sensitivity}, we compare the one-fold
$^2\text{H}(\gamma,n)K^+\Lambda$ d.c.s.\ as calculated with both forms
of the transition amplitude.  Clearly, the RPWIA result and the
on-shell approximation almost coincide for
$|\vn|\lesssim350\,\text{MeV}$.  At large \vn, the
results bifurcate, with the on-shell form of the transition amplitude
giving significantly larger cross sections than the RPWIA.
Accordingly, in phase-space regions with small $|\vn|$,
off-shell ambiguities are absent,
and one can extract information on the on-shell $n(\gamma,K)Y$ amplitude.

\begin{figure}
 \centering
\includegraphics[scale=.59]{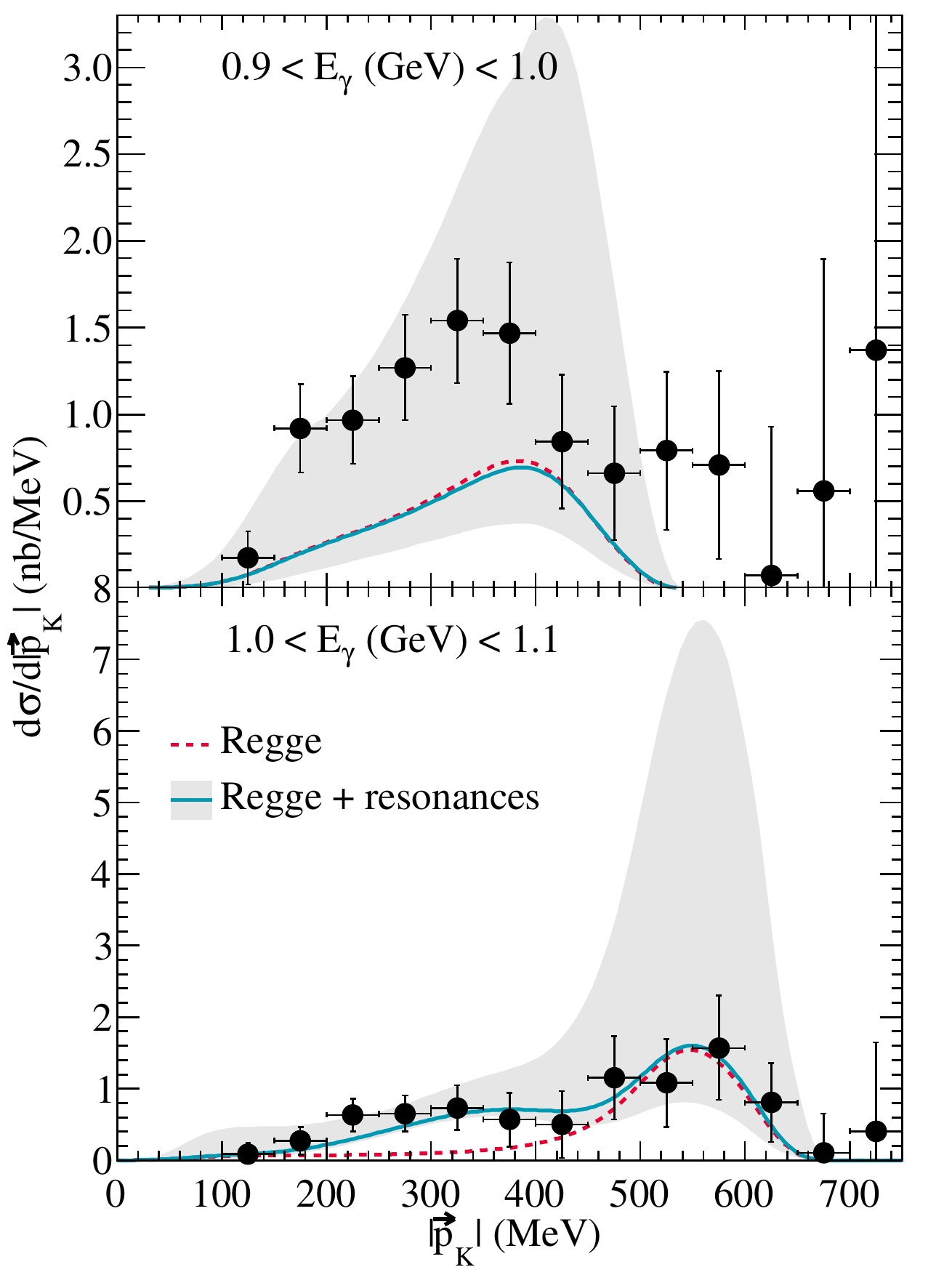} \caption{%
   The semi-inclusive ${^2\text{H}}(\gamma,K^0)YN$ differential cross
   section as a function of the kaon momentum $|\vk|$ integrated over
   $0.9\leq\cos\theta_{K}\leq1$.  The results in the upper
   (lower) panel have been calculated at
   $E_{\gamma}=950\,(1050)\,\text{MeV}$.  The solid~(dashed) line
   shows the result using the full RPR~(Reggeized background)
   amplitude.  The shaded area shows the effect of the uncertainties on the adopted
   helicity amplitudes given in Table~\ref{tab:helamp}. The data are from
   Refs.~\cite{LNSiso4,LNSerratum}.}
\label{fig:TohokuNKS}
\end{figure}

As illustrated earlier, the deuteron density shapes the \KYN cross
section.  The left panel of \figref{sensitivity} shows the $|\vn|$
dependence of the $^2\text{H}(\gamma,n)K^+\Lambda$ d.c.s.\ for various
deuteron wave functions.  For $|\vn|\lesssim 300\,\text{MeV}$ one
obtains nearly indistinguishable results.  This comes as no surprise,
because all $NN$ potentials produce comparable $^3S_1$ waves.  As the
spectator nucleon's momentum rises, the cross-section predictions
start to diverge.  The non-relativistic wave functions of the
Paris~\cite{ParisPot} and Nijmegen~\cite{NijmegenPotential} potentials
and the relativistic Gross-IIB~\cite{GrossWF} wave function generate
very similar predictions.  The cross sections based on the
CD-Bonn~\cite{CDBonnWF} and WJC-1~\cite{GrossWJC} potentials, on the
other hand, differ up to an order of magnitude at high missing
momenta.

To date, the only published \KYN data are from the Laboratory for
Nuclear Science~(LNS) at Tohoku University~\cite{LNSiso4,LNSerratum}.
The semi-inclusive $^2\text{H}(\gamma,K^{0})YN$ cross sections have been
measured in two 100 MeV-wide $E_{\gamma}$ bins close to threshold.

\begin{figure*}
 \centering
\includegraphics[scale=.59]{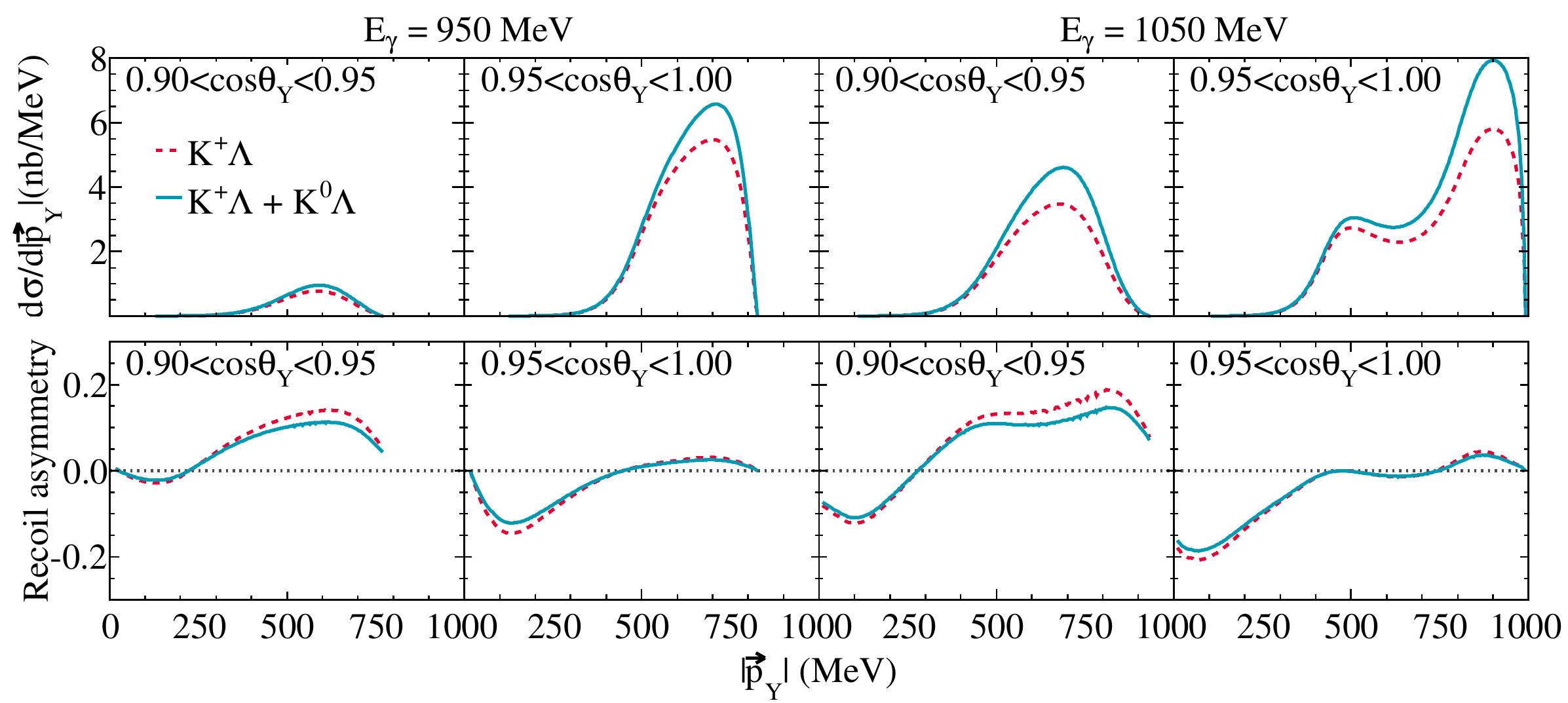} \caption{%
   The semi-inclusive differential cross section~(top row) and recoil
   asymmetry~(bottom row) for $^2\text{H}(\gamma,\Lambda)KN$ as a function
   of the hyperon momentum $|\vy|$ integrated over the $\theta_Y$ bin
   marked in each panel.  The results in the left~(right) panels
   have been calculated at $E_{\gamma}=950\,(1050)\,\text{MeV}$.  The
   solid line shows the result using the RPR amplitude in the RPWIA,
   whereas the dashed curve singles out the
   $^2\text{H}(\gamma,\Lambda)K^+n$ contribution.}
\label{fig:nks2}
\end{figure*}

In \figref{TohokuNKS}, our RPWIA results are compared to the LNS data.
The model calculations are performed at the centre of the $E_{\gamma}$
bin.  As energies close to the $\Sigma$-production threshold are
probed, the cross section has a uni-modal \vk{} distribution
 at $\left< E_{\gamma} \right>=
950\,\text{MeV}$, whereas a second structure arises for the $\left<
  E_{\gamma} \right>= 1050\,\text{MeV}$ bin.  The RPWIA predictions
reproduce the shape of the data and the characteristic quasi-elastic
peaks of semi-inclusive kaon production.  At $\left< E_{\gamma}
\right>= 950\,\text{MeV}$ the strength is underpredicted by roughly a
factor of two. The resonant contributions play an unsubstantial role.
At $\left<E_{\gamma}\right>=1050\,\text{MeV}$, the RPWIA reproduces the measured
magnitude of the cross sections.  The reaction is dominated by the
Reggeized background but the $N^{\star}$ contributions are essential
to reproduce the observed shoulder at $|\vk|\approx300\,\text{MeV}$,
which corresponds to quasi-elastic $\Sigma$ production.

In Section~\ref{sec:elementary}, the error bars on experimental
helicity amplitudes emerged as a chief source of theoretical
uncertainties.  Because both $K^0\Lambda$ and $K^0\Sigma^0$ production
from the neutron contribute to $^2\text{H}(\gamma,K^0)YN$, one can
expect a considerable impact.  The shaded band in \figref{TohokuNKS}
represents the cross sections obtained with the range of coupling
constants of Table~1. At $\left<E_{\gamma}\right>=950\,\text{MeV}$, where the RPR
predictions with the central values of the coupling constants
underestimate the data, the missing strength can be compensated by
including the helicity-amplitude errors.  The errors in the
quasi-elastic $\Lambda$-production peak are large compared to those
for $\Sigma$ production.  This can be understood if one considers the
resonant content of the RPR model for $\Lambda$ production.  The
resonances $P_{13}(1900)$ and $D_{13}(1900)$ play a role and their
photon-helicity couplings have not been determined experimentally.
This forced us to introduce sizable error bars on the ratios of their
coupling constants.

In Ref.~\cite{LNSiso4} it is argued
that the shape of the kaon momentum spectra of the data presented in \figref{TohokuNKS}
is mainly determined
by the angular distribution of the elementary kaon-production cross sections.
The RPR-2007 model, 
which is used as elementary-production amplitude in this work,
has been fitted to $K^+\Lambda$ and $K^+\Sigma^0$ production data at forward kaon angles.
Therefore, for the
deuteron calculations there is some uncertainty stemming from the contributions
of backward kaon angles. We have verified, however, that the semi-inclusive
$^2\text{H}(\gamma,K^0)YN$ differential cross sections
are dominated by the strength from forward angles.

In recent years, dedicated efforts to expand the \KYN database have
been undertaken.  At LNS, new data have been collected with an upgraded
spectrometer~\cite{NKS2procBrian,NKS2procKenta}. Thereby, the hyperon
polarisation becomes accessible~\cite{NKS2procKanda}.  In
\figref{nks2}, RPR-model hyperon-momentum distributions are presented
for the semi-inclusive cross section and recoil asymmetry at LNS
kinematics.  The shape of the cross section is determined by the
momentum of the struck nucleon, and changes as a function of
$E_{\gamma}$ and $\vy$.  The reaction is dominated by $K^+\Lambda$
production.  Our calculations suggest a moderate recoil asymmetry that
changes sign as a function of $E_\gamma$ and $|\vy|$.  For
$\cos\theta_Y\in\left[0.95,1\right]$, the asymmetry is close to zero,
except at low $|\vy|$ where the cross section nearly
vanishes.  For $\cos\theta_Y\in\left[0.9,0.95\right]$, the
asymmetry is mildly negative at small $|\vy|$, and grows to $0.1-0.15$
at quasi-elastic kinematics.  Here, the $K^0\Lambda$
channel has its largest effect, and reduces the size of the recoil
asymmetry.


\section{Conclusions and outlook}
\label{sec:conclusions}

In the proposed RPR framework for strangeness photoproduction, the
analysis of the resonant and non-resonant content of kaon production
is effectively decoupled.  We gauge the predictive power of the
RPR-2007 model, whose parameters are constrained by $p(\gamma,K^+)Y^0$
data, and extend the formalism to reactions with a neutron target and/or a
neutral kaon in the final state.

The RPR-2007 production operator is used in a covariant formalism to
study \KYN reactions.  In the leading RPWIA
contribution to the reaction amplitude, the momentum distribution of
the deuteron emerges as the dominant factor that dictates the angular
and momentum dependence of the cross sections.  Uncertainties related
to the deuteron wave function or off-shell effects are only important
at large missing momenta.  Hence, the elementary amplitude can be most
readily determined based on data obtained at small spectator-nucleon
momenta.

Model predictions for semi-inclusive $K^0$ photoproduction in the
threshold region are compared to experimental results. All predictions
compare favourably to the data.  The incomplete knowledge of helicity
amplitudes induces important uncertainties on the model calculations.

Anticipating new $^2\text{H}(\gamma,\Lambda)KN$ data, we presented predictions for
semi-inclusive $\Lambda$-production cross sections and recoil
asymmetries.  In future work, the quality of our formalism will
benefit from the new RPR-model analysis of the world's \KpL
data~\cite{RPRprl,RPR2011}.  In addition, inclusion of the different
rescattering contributions will allow to study exclusive kaon
photoproduction at a more detailed level, and single out phase-space
regions where the elusive hyperon-nucleon potential can be
investigated.


\section*{Acknowledgements} 
This work was supported by the Research Foundation -- Flanders (FWO) and the research council of Ghent University.
The calculations were carried out using the Stevin Supercomputer Infrastructure at Ghent University,
funded by Ghent University, the Hercules Foundation and the Flemish Government -- department EWI.


\bibliographystyle{model1-num-names}
\bibliography{bibliography}

\begin{thebibliography}{44}
\expandafter\ifx\csname natexlab\endcsname\relax\def\natexlab#1{#1}\fi
\providecommand{\bibinfo}[2]{#2}
\ifx\xfnm\relax \def\xfnm[#1]{\unskip,\space#1}\fi
\bibitem[{Klempt and Richard(2010)}]{ResonanceReview}
\bibinfo{author}{E.~Klempt}, \bibinfo{author}{J.-M. Richard},
\newblock \bibinfo{journal}{Rev.Mod.Phys.} \bibinfo{volume}{82}
  (\bibinfo{year}{2010}) \bibinfo{pages}{1095}.
\bibitem[{Burkert and Lee(2004)}]{ReviewExperiments}
\bibinfo{author}{V.~Burkert}, \bibinfo{author}{T.~Lee},
\newblock \bibinfo{journal}{Int.J.Mod.Phys.} \bibinfo{volume}{E13}
  (\bibinfo{year}{2004}) \bibinfo{pages}{1035}.
\bibitem[{Guidal et~al.(1997)Guidal, Laget, and
  Vanderhaeghen}]{GuidalPhotoProdKandPi}
\bibinfo{author}{M.~Guidal}, \bibinfo{author}{J.~M. Laget},
  \bibinfo{author}{M.~Vanderhaeghen},
\newblock \bibinfo{journal}{Nucl.Phys.} \bibinfo{volume}{A627}
  (\bibinfo{year}{1997}) \bibinfo{pages}{645}.
\bibitem[{De~Cruz et~al.(2010)De~Cruz, Ireland, Vancraeyveld, and
  Ryckebusch}]{RPRbayes}
\bibinfo{author}{L.~De~Cruz}, \bibinfo{author}{D.~G. Ireland},
  \bibinfo{author}{P.~Vancraeyveld}, \bibinfo{author}{J.~Ryckebusch},
\newblock \bibinfo{journal}{Phys.Lett.} \bibinfo{volume}{B694}
  (\bibinfo{year}{2010}) \bibinfo{pages}{33}.
\bibitem[{Corthals et~al.(2006)Corthals, Ryckebusch, and
  Van~Cauteren}]{RPRlambda}
\bibinfo{author}{T.~Corthals}, \bibinfo{author}{J.~Ryckebusch},
  \bibinfo{author}{T.~Van~Cauteren},
\newblock \bibinfo{journal}{Phys.Rev.} \bibinfo{volume}{C73}
  (\bibinfo{year}{2006}) \bibinfo{pages}{045207}.
\bibitem[{Corthals et~al.(2007{\natexlab{a}})Corthals, Ireland, Van~Cauteren,
  and Ryckebusch}]{RPRsigma}
\bibinfo{author}{T.~Corthals}, \bibinfo{author}{D.~G. Ireland},
  \bibinfo{author}{T.~Van~Cauteren}, \bibinfo{author}{J.~Ryckebusch},
\newblock \bibinfo{journal}{Phys.Rev.} \bibinfo{volume}{C75}
  (\bibinfo{year}{2007}{\natexlab{a}}) \bibinfo{pages}{045204}.
\bibitem[{Corthals et~al.(2007{\natexlab{b}})Corthals, Van~Cauteren,
  Vancraeyveld, Ryckebusch, and Ireland}]{RPRelectro}
\bibinfo{author}{T.~Corthals}, \bibinfo{author}{T.~Van~Cauteren},
  \bibinfo{author}{P.~Vancraeyveld}, \bibinfo{author}{J.~Ryckebusch},
  \bibinfo{author}{D.~G. Ireland},
\newblock \bibinfo{journal}{Phys.Lett.} \bibinfo{volume}{B656}
  (\bibinfo{year}{2007}{\natexlab{b}}) \bibinfo{pages}{186}.
\bibitem[{Vancraeyveld et~al.(2009)Vancraeyveld, De~Cruz, Ryckebusch, and
  Van~Cauteren}]{RPRneutron}
\bibinfo{author}{P.~Vancraeyveld}, \bibinfo{author}{L.~De~Cruz},
  \bibinfo{author}{J.~Ryckebusch}, \bibinfo{author}{T.~Van~Cauteren},
\newblock \bibinfo{journal}{Phys.Lett.} \bibinfo{volume}{B681}
  (\bibinfo{year}{2009}) \bibinfo{pages}{428}.
\bibitem[{Vancraeyveld(2011)}]{PhdPieter}
\bibinfo{author}{P.~Vancraeyveld}, Ph.D. thesis, Ghent University,
  \bibinfo{year}{2011}.
  \bibinfo{note}{{http://inwpent5.ugent.be/Publication/phd/phdpietervan.pdf}}.
\bibitem[{Merten et~al.(2002)Merten, L{\"o}ring, Kretzschmar, Metsch, and
  Petry}]{BonnEMff}
\bibinfo{author}{D.~Merten}, \bibinfo{author}{U.~L{\"o}ring},
  \bibinfo{author}{K.~Kretzschmar}, \bibinfo{author}{B.~Metsch},
  \bibinfo{author}{H.~R. Petry},
\newblock \bibinfo{journal}{Eur.Phys.J.} \bibinfo{volume}{A14}
  (\bibinfo{year}{2002}) \bibinfo{pages}{477}.
\bibitem[{Arndt et~al.(1996)Arndt, Strakovsky, and Workman}]{SAID96}
\bibinfo{author}{R.~A. Arndt}, \bibinfo{author}{I.~I. Strakovsky},
  \bibinfo{author}{R.~L. Workman},
\newblock \bibinfo{journal}{Phys.Rev.} \bibinfo{volume}{C53}
  (\bibinfo{year}{1996}) \bibinfo{pages}{430}.
\bibitem[{Kohri et~al.(2006)}]{LEPSiso6}
\bibinfo{author}{H.~Kohri}, et~al.,
\newblock \bibinfo{journal}{Phys.Rev.Lett} \bibinfo{volume}{97}
  (\bibinfo{year}{2006}) \bibinfo{pages}{082003}.
\bibitem[{Anefalos~Pereira et~al.(2010)}]{AnefalosIso6}
\bibinfo{author}{S.~Anefalos~Pereira}, et~al.,
\newblock \bibinfo{journal}{Phys.Lett.} \bibinfo{volume}{B688}
  (\bibinfo{year}{2010}) \bibinfo{pages}{289}.
\bibitem[{Beringer et~al.(2012)}]{pdg}
\bibinfo{author}{J.~Beringer}, et~al.,
\newblock \bibinfo{journal}{Phys.Rev.} \bibinfo{volume}{D86}
  (\bibinfo{year}{2012}) \bibinfo{pages}{010001}.
\bibitem[{Singer and Miller(1986)}]{SingerMiller}
\bibinfo{author}{P.~Singer}, \bibinfo{author}{G.~A. Miller},
\newblock \bibinfo{journal}{Phys.Rev.} \bibinfo{volume}{D33}
  (\bibinfo{year}{1986}) \bibinfo{pages}{141}.
\bibitem[{Van~Cauteren et~al.(2005)Van~Cauteren, Ryckebusch, Metsch, and
  Petry}]{VanCauteren:2005sm}
\bibinfo{author}{T.~Van~Cauteren}, \bibinfo{author}{J.~Ryckebusch},
  \bibinfo{author}{B.~Metsch}, \bibinfo{author}{H.-R. Petry},
\newblock \bibinfo{journal}{Eur.Phys.J.} \bibinfo{volume}{A26}
  (\bibinfo{year}{2005}) \bibinfo{pages}{339}.
\bibitem[{Yamamura et~al.(2000)Yamamura, Miyagawa, Mart, Bennhold, and
  Gloeckle}]{Yamamura}
\bibinfo{author}{H.~Yamamura}, \bibinfo{author}{K.~Miyagawa},
  \bibinfo{author}{T.~Mart}, \bibinfo{author}{C.~Bennhold},
  \bibinfo{author}{W.~Gloeckle},
\newblock \bibinfo{journal}{Phys.Rev.} \bibinfo{volume}{C61}
  (\bibinfo{year}{2000}) \bibinfo{pages}{014001}.
\bibitem[{Miyagawa et~al.(2006)Miyagawa, Mart, Bennhold, and
  Gl{\"o}ckle}]{Miyagawa2006}
\bibinfo{author}{K.~Miyagawa}, \bibinfo{author}{T.~Mart},
  \bibinfo{author}{C.~Bennhold}, \bibinfo{author}{W.~Gl{\"o}ckle},
\newblock \bibinfo{journal}{Phys.Rev.} \bibinfo{volume}{C74}
  (\bibinfo{year}{2006}) \bibinfo{pages}{034002}.
\bibitem[{Salam and Arenh{\"o}vel(2004)}]{Salam2004}
\bibinfo{author}{A.~Salam}, \bibinfo{author}{H.~Arenh{\"o}vel},
\newblock \bibinfo{journal}{Phys.Rev.} \bibinfo{volume}{C70}
  (\bibinfo{year}{2004}) \bibinfo{pages}{044008}.
\bibitem[{Salam et~al.(2006)Salam, Miyagawa, Mart, Bennhold, and
  Gl{\"o}ckle}]{Salam2006}
\bibinfo{author}{A.~Salam}, \bibinfo{author}{K.~Miyagawa},
  \bibinfo{author}{T.~Mart}, \bibinfo{author}{C.~Bennhold},
  \bibinfo{author}{W.~Gl{\"o}ckle},
\newblock \bibinfo{journal}{Phys.Rev.} \bibinfo{volume}{C74}
  (\bibinfo{year}{2006}) \bibinfo{pages}{044004}.
\bibitem[{Salam et~al.(2009)Salam, Mart, and Miyagawa}]{Salam2009}
\bibinfo{author}{A.~Salam}, \bibinfo{author}{T.~Mart},
  \bibinfo{author}{K.~Miyagawa},
\newblock \bibinfo{journal}{Mod.Phys.Lett.} \bibinfo{volume}{A24}
  (\bibinfo{year}{2009}) \bibinfo{pages}{968}.
\bibitem[{Kerbikov(2001)}]{Kerbikov2000}
\bibinfo{author}{B.~O. Kerbikov},
\newblock \bibinfo{journal}{Phys.Atom.Nucl.} \bibinfo{volume}{64}
  (\bibinfo{year}{2001}) \bibinfo{pages}{1835}.
\bibitem[{Maxwell(2004{\natexlab{a}})}]{Maxwell2004a}
\bibinfo{author}{O.~V. Maxwell},
\newblock \bibinfo{journal}{Phys.Rev.} \bibinfo{volume}{C69}
  (\bibinfo{year}{2004}{\natexlab{a}}) \bibinfo{pages}{034605}.
\bibitem[{Maxwell(2004{\natexlab{b}})}]{Maxwell2004}
\bibinfo{author}{O.~V. Maxwell},
\newblock \bibinfo{journal}{Phys.Rev.} \bibinfo{volume}{C70}
  (\bibinfo{year}{2004}{\natexlab{b}}) \bibinfo{pages}{044612}.
\bibitem[{Bydzovsky(2010)}]{Bydzovsky2010a}
\bibinfo{author}{P.~Bydzovsky},
\newblock \bibinfo{journal}{Int.J.Mod.Phys.} \bibinfo{volume}{E19}
  (\bibinfo{year}{2010}) \bibinfo{pages}{2369}.
\bibitem[{Bydzovsky and Sotona(2010)}]{Bydzovsky2010b}
\bibinfo{author}{P.~Bydzovsky}, \bibinfo{author}{M.~Sotona},
\newblock \bibinfo{journal}{Nucl.Phys.} \bibinfo{volume}{A835}
  (\bibinfo{year}{2010}) \bibinfo{pages}{246}.
\bibitem[{Gasparyan et~al.(2007)Gasparyan, Haidenbauer, Hanhart, and
  Miyagawa}]{Gasparyan2007}
\bibinfo{author}{A.~Gasparyan}, \bibinfo{author}{J.~Haidenbauer},
  \bibinfo{author}{C.~Hanhart}, \bibinfo{author}{K.~Miyagawa},
\newblock \bibinfo{journal}{Eur.Phys.J.} \bibinfo{volume}{A32}
  (\bibinfo{year}{2007}) \bibinfo{pages}{61}.
\bibitem[{Laget(2006)}]{Laget2006}
\bibinfo{author}{J.-M. Laget},
\newblock \bibinfo{journal}{Phys.Rev.} \bibinfo{volume}{C73}
  (\bibinfo{year}{2006}) \bibinfo{pages}{044003}.
\bibitem[{Laget(2007)}]{Laget2006pentaquark}
\bibinfo{author}{J.-M. Laget},
\newblock \bibinfo{journal}{Phys.Rev.} \bibinfo{volume}{C75}
  (\bibinfo{year}{2007}) \bibinfo{pages}{014002}.
\bibitem[{Blankenbecler and Cook(1960)}]{BlankenbeclerCook}
\bibinfo{author}{R.~Blankenbecler}, \bibinfo{author}{L.~F. Cook},
\newblock \bibinfo{journal}{Phys.Rev.} \bibinfo{volume}{119}
  (\bibinfo{year}{1960}) \bibinfo{pages}{1745}.
\bibitem[{Buck and Gross(1979)}]{BuckGross}
\bibinfo{author}{W.~W. Buck}, \bibinfo{author}{F.~Gross},
\newblock \bibinfo{journal}{Phys.Rev.} \bibinfo{volume}{D20}
  (\bibinfo{year}{1979}) \bibinfo{pages}{2361}.
\bibitem[{Laget(1981)}]{LagetPhysRept}
\bibinfo{author}{J.-M. Laget},
\newblock \bibinfo{journal}{Phys.Rept.} \bibinfo{volume}{69}
  (\bibinfo{year}{1981}) \bibinfo{pages}{1--84}.
\bibitem[{Machleidt(2001)}]{CDBonnWF}
\bibinfo{author}{R.~Machleidt},
\newblock \bibinfo{journal}{Phys.Rev.} \bibinfo{volume}{C63}
  (\bibinfo{year}{2001}) \bibinfo{pages}{024001}.
\bibitem[{Lacombe et~al.(1980)}]{ParisPot}
\bibinfo{author}{M.~Lacombe}, et~al.,
\newblock \bibinfo{journal}{Phys.Rev.} \bibinfo{volume}{C21}
  (\bibinfo{year}{1980}) \bibinfo{pages}{861}.
\bibitem[{Stoks et~al.(1994)Stoks, Klomp, Terheggen, and
  de~Swart}]{NijmegenPotential}
\bibinfo{author}{V.~G.~J. Stoks}, \bibinfo{author}{R.~A.~M. Klomp},
  \bibinfo{author}{C.~P.~F. Terheggen}, \bibinfo{author}{J.~J. de~Swart},
\newblock \bibinfo{journal}{Phys.Rev.} \bibinfo{volume}{C49}
  (\bibinfo{year}{1994}) \bibinfo{pages}{2950}.
\bibitem[{Gross et~al.(1992)Gross, Van~Orden, and Holinde}]{GrossWF}
\bibinfo{author}{F.~Gross}, \bibinfo{author}{J.~W. Van~Orden},
  \bibinfo{author}{K.~Holinde},
\newblock \bibinfo{journal}{Phys.Rev.} \bibinfo{volume}{C45}
  (\bibinfo{year}{1992}) \bibinfo{pages}{2094--2132}.
\bibitem[{Gross and Stadler(2008)}]{GrossWJC}
\bibinfo{author}{F.~Gross}, \bibinfo{author}{A.~Stadler},
\newblock \bibinfo{journal}{Phys.Rev.} \bibinfo{volume}{C78}
  (\bibinfo{year}{2008}) \bibinfo{pages}{014005}.
\bibitem[{Tsukada et~al.(2008)}]{LNSiso4}
\bibinfo{author}{K.~Tsukada}, et~al.,
\newblock \bibinfo{journal}{Phys.Rev.} \bibinfo{volume}{C78}
  (\bibinfo{year}{2008}) \bibinfo{pages}{014001}.
\bibitem[{Tsukada et~al.(2011)}]{LNSerratum}
\bibinfo{author}{K.~Tsukada}, et~al.,
\newblock \bibinfo{journal}{Phys.Rev.} \bibinfo{volume}{C83}
  (\bibinfo{year}{2011}) \bibinfo{pages}{039904}.
\bibitem[{Beckford et~al.(2011)}]{NKS2procBrian}
\bibinfo{author}{B.~Beckford}, et~al.,
\newblock \bibinfo{journal}{AIP Conf.Proc.} \bibinfo{volume}{1388}
  (\bibinfo{year}{2011}) \bibinfo{pages}{280}.
\bibitem[{{Futatsukawa} et~al.(2012)}]{NKS2procKenta}
\bibinfo{author}{K.~{Futatsukawa}}, et~al.,
\newblock \bibinfo{journal}{EPJ Web Conf.} \bibinfo{volume}{20}
  (\bibinfo{year}{2012}) \bibinfo{pages}{02005}.
\bibitem[{Kanda et~al.(2010)}]{NKS2procKanda}
\bibinfo{author}{H.~Kanda}, et~al.,
\newblock \bibinfo{journal}{Nucl.Phys.} \bibinfo{volume}{A835}
  (\bibinfo{year}{2010}) \bibinfo{pages}{317}.
\bibitem[{De~Cruz et~al.(2012{\natexlab{a}})De~Cruz, Vrancx, Vancraeyveld, and
  Ryckebusch}]{RPRprl}
\bibinfo{author}{L.~De~Cruz}, \bibinfo{author}{T.~Vrancx},
  \bibinfo{author}{P.~Vancraeyveld}, \bibinfo{author}{J.~Ryckebusch},
\newblock \bibinfo{journal}{Phys.Rev.Lett.} \bibinfo{volume}{108}
  (\bibinfo{year}{2012}{\natexlab{a}}) \bibinfo{pages}{182002}.
\bibitem[{De~Cruz et~al.(2012{\natexlab{b}})De~Cruz, Ryckebusch, Vrancx, and
  Vancraeyveld}]{RPR2011}
\bibinfo{author}{L.~De~Cruz}, \bibinfo{author}{J.~Ryckebusch},
  \bibinfo{author}{T.~Vrancx}, \bibinfo{author}{P.~Vancraeyveld},
\newblock \bibinfo{journal}{Phys.Rev.} \bibinfo{volume}{C86}
  (\bibinfo{year}{2012}{\natexlab{b}}) \bibinfo{pages}{015212}.

\end{thebibliography}

\end{document}